\documentclass[11pt,prd,onecolumn,amsfonts,amssymb,amsmath,showkeys,nofootinbib]{revtex4-1}
\usepackage[english]{babel}
\usepackage[utf8]{inputenc}
\usepackage[colorinlistoftodos, color=green!40, prependcaption]{todonotes}
\usepackage{amsthm}
\usepackage{mathtools}
\usepackage{physics}
\usepackage{xcolor}
\usepackage{graphicx}
\usepackage{adjustbox}
\usepackage{placeins}
\usepackage[T1]{fontenc}
\usepackage{lipsum}
\usepackage{csquotes}
\usepackage[pdftex, pdftitle={Article}, pdfauthor={Author}]{hyperref} % For hyperlinks in the PDF
\usepackage{multirow}
\usepackage{array}
\begin{document}

\begin{flushright}
CERN-TH-2021-088\vspace*{0.5cm}
\end{flushright}

\title{Discriminating same-mass Neutron Stars and Black Holes gravitational wave-forms}

\author{J.-F. Coupechoux$^{1}$}
\email{j-f.coupechoux@ipnl.in2p3.fr}
\author{A. Arbey$^{1,2,3}$}
\author{R. Chierici$^1$}
\author{H. Hansen$^1$}
\author{J. Margueron$^1$}
\author{V. Sordini$^1$}
   
    \affiliation{$^1$Univ Lyon, Univ Claude Bernard Lyon 1, CNRS/IN2P3,\\
    Institut de Physique des 2 Infinis de Lyon, UMR 5822, 69622 Villeurbanne, France\vspace*{0.1cm}\\
    $^2$Theoretical Physics Department, CERN, CH-1211 Geneva 23, Switzerland\vspace*{0.1cm}\\
    $^3$Institut Universitaire de France, 103 boulevard Saint-Michel, 75005 Paris, France}
    
\date{\today} % Leave empty to omit a date

\begin{abstract}
Gravitational wave-forms from coalescences of binary black hole systems and binary neutron star systems with low tidal effects can hardly be distinguished if the two systems have similar masses. In the absence of discriminating power based on the gravitational wave-forms, the classification of sources into binary neutron stars, binary black holes and mixed systems containing a black hole and a neutron star can only be unambiguous when assuming the standard model of stellar evolution and using the fact that there exists a mass gap between neutron stars and black holes. This approach is however limited by its own assumptions: for instance the $2.6$ solar mass object detected in the GW190814 event remains unclassified, and models of new physics can introduce new compact objects, like primordial black holes, which may have masses in the same range as neutron stars. Then, without an electromagnetic counterpart (kilonova), classifying mergers of compact objects without mass gap criteria remains a difficult task, unless the source is close enough. In what follows we investigate a procedure to discriminate a model between binary neutron star merger and primordial binary black hole merger by using a Bayes factor in simulated wave-forms that we superimpose to realistic detector noise. 

%In what follows, we analyze the impact of the source distance, and eventually tidal deformability, on the possibility to distinguish between BBH, BNS and BHNS mergers.

\end{abstract}


\maketitle

\section{Introduction}

Gravitational wave astronomy has entered a new era, characterized by the detection of a plethora of coalescing compact binary objects from the recent LIGO and Virgo runs~\cite{LIGOScientific:2018mvr, Abbott:2020niy}, with a large range of masses for the individual compact objects. In particular, the GW190814 event \cite{Abbott:2020khf} corresponds to a merger involving a compact object with a $2.6$ solar masses. Such a mass falls typically in the intermediate range between the known black hole masses and neutron star masses~\cite{Rezzolla_2018, LIGOScientific:2019eut, Pejcha:2014wda}, therefore this object can be either the most massive neutron star or the lightest black hole ever detected. In terms of gravitational wave emission, the main difference to be expected is related to the condensed matter effect of neutron stars. Unfortunately, the detector sensitivities are currently too low to observe the impact of matter effect, and in absence of electromagnetic counterpart to determine the nature of this compact object. The mechanisms of formation of a black hole or a neutron star of $2.6$ solar masses remains unclear and an interesting idea is to consider that this object is a primordial black hole (PBHs)~\cite{Byrnes:2018clq}. 

Contrary to stellar black holes which are produced by supernovae, PBHs may have been produced in the early Universe, during e.g. a phase transition. Their size is limited by the Hubble scale, which is related to the cosmological time, providing a link between the maximum PBH mass and the epoch of formation \cite{Carr:2020xqk}, and in practice PBHs can theoretically have a mass between the Planck mass and millions of solar masses. Because of Hawking evaporation, light PBHs lose mass under the form of radiation, and it has been shown that PBHs with masses below $10^{15}$ grams would have already evaporated since their formation \cite{Carr:2020xqk}, but heavier PBHs have sufficiently low evaporation rates to still remain. Because of this, PBHs can constitute the whole or a large fraction of dark matter, and they are usually considered as good candidates for dark matter~\cite{Carr:2020xqk}. The GW190814 event is not the only event to contain a compact object of unknown nature because of a too low resolution of detectors ; the question about the nature of compact objects can also be asked for the GW190425 event~\cite{Abbott:2020uma}.

In this article, we study the differences between gravitational wave-forms (GWs) produced by the coalescence of binary neutron stars (BNS), binary black holes (BBH) and mixed systems containing a black hole and a neutron star (BHNS). The shapes of these gravitational wave-forms are rather similar when the objects involved in the merger have similar masses. The main goal of this paper is to understand under which circumstances and conditions it is possible to discriminate  between BBH and BNS wave-forms by making use of injected signals with realistic noise profiles. In section~II, BBH, BHNS and BNS templates will be compared using the match which is equivalent to the overlap maximized over time and phase, and the odds number will be introduced to compare two competing models. In section~III, we will study the possibility of misinterpreting the results when injecting a PBH wave-form into advanced detector noise and interpreting the data in the context of BNS mergers. We will also explore the opposite problem, when BNSs are interpreted as BBHs, and we will conclude in Section IV. 

%%%%%%%%%%%%%%%%%%%%%%%%%%%%%%%%%%%%%%%%%%%%%

\section{Model selection}

\subsection{Degeneracy between BNS, BHNS and BBH wave-forms}

A first approach to compare two GWs is to use the convolution product of the two GWs in the time domain which is defined as the integral of the product of two GWs after one of them has been reversed, shifted and extended by zero values. The discriminating parameter to vary between BNS, BHNS and BBH wave-forms is the tidal deformability characterizing the matter effect~\cite{Flanagan:2007ix, Hinderer:2007mb, Hinderer2009} which is only defined for neutron stars. Without mass gap hypothesis, it is this parameter that will play a central role in characterizing the nature of compact objects. For example, the convolution product of the wave-forms between a BBH and a BNS normalized with the convolution product of BBH with itself is $0.937$ for tidal deformabilities equal to $\Lambda_1=\Lambda_2=600$ and $0.997$ for $\Lambda_1=\Lambda_2=0$.

\begin{figure}[t!]
\centering
\includegraphics[width=12cm]{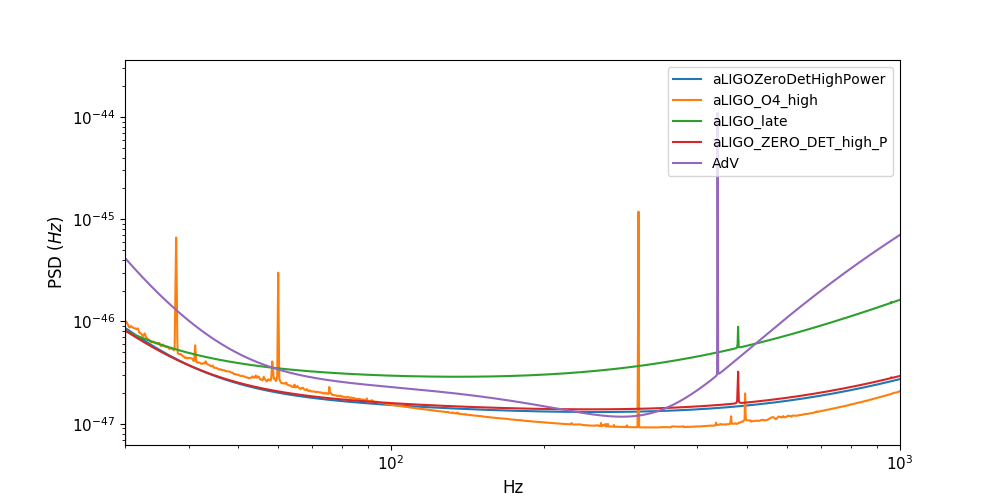}
\caption{Power spectral density of noise for the advanced LIGO and Virgo detector configurations.}
\label{fig:noise}
\end{figure}

Following the article~\cite{Lindblom:2008cm}, the approach to compare two wave-forms taking into account the sensitivity of the detectors, is based on the noise-weighted inner product between two wave-forms $h_1$ and $h_2$ defined by:
\begin{equation}\label{eq:innerproduct}
    \langle h_1 | h_2 \rangle = 4 \Re \int_{f_{min}}^{f_{max}} \frac{h_1(f)h_2^*(f)}{S_n(f)}\,,
\end{equation}
where $S_n(f)$ is the power spectral density, that encodes the frequency-dependent sensitivity of a detector~\cite{Cutler:1994ys}. To calculate such scalar product, we take the advanced LIGO design sensitivity given by \textit{aLIGOZeroDetHighPower}~\cite{alex_nitz_2021_4556907}, plotted in Figure~\ref{fig:noise}. The separation between two wave-forms $h_1$ and $h_2$ can be related to their match, which is defined by the overlap maximized on the coalescence time and the coalescence phase:
\begin{equation}
M(h_1,h_2) = \max_{\Delta t, \Delta \phi} \frac{\langle h_1 | h_2 \rangle }{\sqrt{\langle h_1 | h_1 \rangle }\sqrt{\langle h_2 | h_2 \rangle }} \,.
\label{eq:match}
\end{equation}
In gravitational wave searches for Compact Binary Coalescences (CBC), the LIGO/Virgo Collaboration constructs template banks so that the match is at least $0.97$ between the two closest wave-forms \cite{Babak:2006ty}. To measure a difference between a neutron star and a black hole via their gravitational wave-forms without taking into account the masses of the compact objects, it is necessary to detect a significant matter effect affecting the wave-form through the tidal deformability. To give an example of the difficulty of such a task, the match between a BBH and a BNS with $\Lambda_1=\Lambda_2=600$ with a chirp mass of $1.44$~$\text{M}_\odot$ and a mass ratio of $0.9$ is higher than $0.97$. It requires at least $\Lambda_1=\Lambda_2=800$ to have a match lower than $0.97$.  

\begin{figure}[t!]
\centering
\includegraphics[width=12cm]{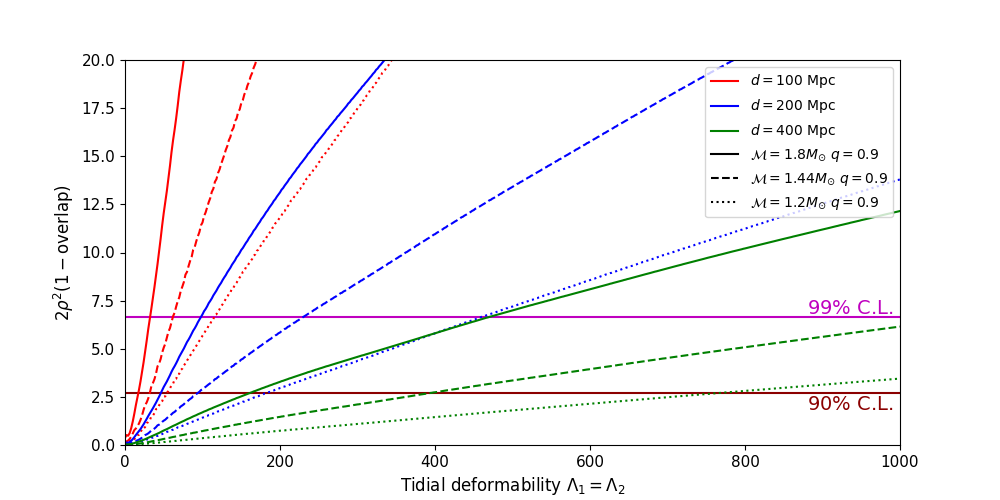}
\caption{Comparison between BBH and BNS wave-forms. The curves show the left-hand side of Eq.~\eqref{comparison} as a function of the tidal deformabilities of the neutron stars. The horizontal lines show the value of $\chi^2_{k}$ for different confidence level thresholds : if a point is below the horizontal lines, the two mergers are too similar to be distinguished at the given confidence level. Different types of curves (solid, dashed and dotted) of the same color indicate different chirp masses and different types of colors indicate different distances. In particular, ($\mathcal{M}=1.2M_{\odot}$, $q=0.9$) corresponds to ($m_1=1.45M_{\odot}$, $m_2=1.31M_{\odot}$), ($\mathcal{M}=1.44M_{\odot}$, $q=0.9$) to ($m_1=1.74M_{\odot}$, $m_2=1.57M_{\odot}$) and ($\mathcal{M}=1.8M_{\odot}$, $q=0.9$) to ($m_1=2.18M_{\odot}$, $m_2=1.96M_{\odot}$)}
\label{fig:BBH_BNS}
\end{figure}

When a signal is detected, given a model $\mathcal{M}_A$, the marginalized posterior probability density function of all unknown parameters $\theta$ is typically computed by using a Bayesian analysis. The mean of the parameters $\theta$ is noted $\langle \theta \rangle$. Following the Eq. (18) of~\cite{Baird:2012cu}, the confidence region at a given probability $p$ is the set of points that verifies the following condition:
\begin{equation}
    2 \rho^2 \left[ 1 - M(h(\theta),h(\langle \theta \rangle)) \right] \leq \chi_k^2(1-p)\,,
    \label{match}
\end{equation}
where $\rho$ is the signal to noise ratio (SNR) and $\chi^2_k(1-p)$ is the chi-squared, in the case of $k$ degrees of freedom. We can use this relation to determine the regions in the parameter space in which two different wave-forms can be distinguished. For a given SNR, if spins are neglected, BNS templates have only $4$ parameters: the chirp mass $\mathcal{M}$, the mass ratio $q$ and the two tidal deformabilities $\Lambda_1$, $\Lambda_2$ because the match is maximized over phase, time and distance. Since we want to measure matter effects to distinguish between a neutron star and a black hole, we consider only the tidal deformabilities as free parameters and fix the others. In addition, we make the simplifying assumption that $\Lambda_1=\Lambda_2$. We write Eq.~\eqref{match} for the case where $h(\langle \theta \rangle)$ is a BBH wave-form and for the case where $h(\theta )$ is a BNS one with $\Lambda_1=\Lambda_2$. It follows that BBH and BNS can be distinguished (and the nature of the compact object can be determined) if:
\begin{equation}
    2 \rho^2 \left[ 1 - M(h_{\text{BNS}},h_{\text{BBH}}) \right] \geq \chi_k^2(1-p)\,,
    \label{comparison}
\end{equation}
with $ \chi_k^2(1-p) = 2.71$ at  $90\%$ C.L., which corresponds to one degree of freedom since the tidal deformability is the only free parameter ($\Lambda_1=\Lambda_2$). The left-hand side of Eq.~(\ref{comparison}) depends on the fixed parameters but also on the luminosity distance $d$ through the SNR: $\rho \simeq \langle h_{\text{BBH}}| h_{\text{BBH}}\rangle^{1/2} \simeq \langle h_{\text{BNS}}| h_{\text{BNS}}\rangle^{1/2} $ which is proportional to $1/d$. Figure~\ref{fig:BBH_BNS} shows $2 \rho^2 \left[ 1 - M(h_{\text{BNS}},h_{\text{BBH}}) \right]$ as a function of the deformability of neutron stars.  The horizontal lines correspond to the $90$ and $99\%$ C.L. limits: a point above one of these lines means that BBH and BNS mergers can be distinguished at more than 90\% of 99\% C.L. The wave-forms are calculated using {\tt IMRPhenomPv2}~\cite{Husa:2015iqa, Khan:2015jqa, Khan:2018fmp} for BBH and {\tt IMRPhenomPv2\_NRTidal}~\cite{Dietrich:2017aum, Dietrich:2018uni, Dietrich:2019kaq} for BNS. As expected, the ability of distinguishing BBH and BNS increases with the tidal deformabilities and the chirp mass, and decreases with the distances. For example, for a distance of $d=200$ Mpc, with $\mathcal{M}=1.44$ $M_{\odot}$ and $q=0.9$, the nature of the compact objects can be determined if the tidal deformabilities are higher than $200$, while  for a distance of $400$ Mpc even tidal deformabilities of $1000$ are not enough to make the BBH and BNS wave-forms distinguishable at $99\%$ C.L.

\begin{figure}[t!]
\centering
\includegraphics[width=12cm]{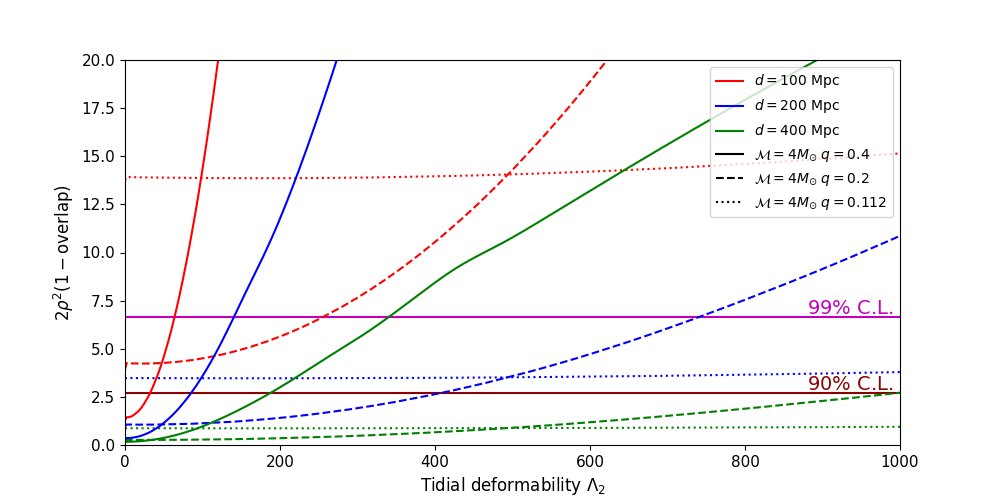}
\caption{Same as Figure~\ref{fig:BBH_BNS} for BBH and BHNS templates. ($\mathcal{M}=4M_{\odot}$, $q=0.112$) corresponds to ($m_1=15.2M_{\odot}$, $m_2=1.7M_{\odot}$), ($\mathcal{M}=4M_{\odot}$, $q=0.2$) to ($m_1=10.9M_{\odot}$, $m_2=2.2M_{\odot}$) and ($\mathcal{M}=4M_{\odot}$, $q=0.4$) to ($m_1=7.4M_{\odot}$, $m_2=3.0M_{\odot}$).}
\label{fig:BBH_BHNS}
\end{figure}

We performed a similar study for the comparison of BBH and BHNS wave-forms, to determine the conditions under which it is possible to distinguish the nature of the compact object coalescing with the black hole. For this case, the BBH wave-forms are generated with {\tt IMRPhenomPv3HM}~\cite{Khan:2019kot}, because higher-order modes are important for such asymmetric systems, and {\tt IMRPhenomNSBH}~\cite{Thompson:2020nei} is used for BHNS. The results are shown in Figure~\ref{fig:BBH_BHNS}, for different asymmetric systems. For a small mass ratio of $0.112$, the overlap between BBH and BHNS wave-forms does not really depend on the tidal deformability of the second compact object. Indeed, in such asymmetric configurations, the coalescence evolution is driven by the most massive object: the smallest object is absorbed by the black hole and its tidal deformability has a negligible effect. In this case, for a chirp mass of $4$ $M_{\odot}$, the nature of the companion can be determined only for a system at low distance (less than $200$~Mpc).  The more symmetric the system, the more important the tidal deformability is to distinguish BBH and BHNS wave-forms.    

\begin{figure}[h!]
\centering
\includegraphics[width=12cm]{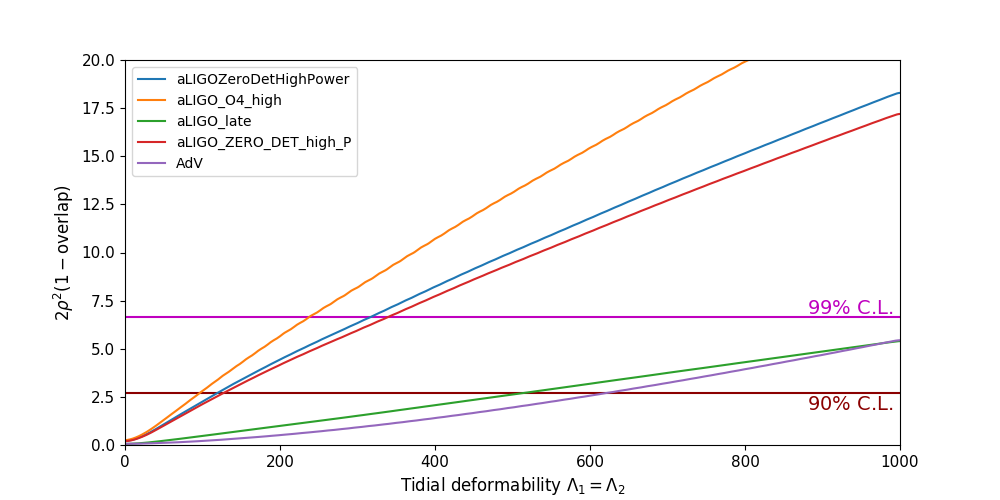}
\caption{Comparison between BBH and BNS mergers, assuming the different power spectral densities shown in Figure~\ref{fig:noise}. The chirp mass is fixed at $1.44$ $M_{\odot}$, the mass ratio at $0.9$ and the distance at $200$ Mpc.\label{fig:noise_BBH_BNS}}
\end{figure}

The results shown in Figures~\ref{fig:BBH_BNS} and \ref{fig:BBH_BHNS} depend strongly on the detector sensitivity. Indeed, any change in the power spectral density (PSD) $S_n(f)$ directly affects the noise-weighted inner product between two wave-forms (Eq.~(\ref{eq:innerproduct})). Some examples of PSD are drawn in Figure~\ref{fig:noise}. Their impact on the ability to distinguish between BBH and BNS templates and to determine the nature of compact objects is shown in Figure~\ref{fig:noise_BBH_BNS} for the system defined by $\mathcal{M}=1.44$~$M_{\odot}$, $q=0.9$ and $d=200$~Mpc.

\subsection{The odds number}

The Bayes' theorem links the posterior distribution to the likelihood, the prior and the evidence~\cite{Thrane:2018qnx}:
\begin{equation}
p(\theta | d,\mathcal{M}_A) = \frac{\pi(\theta | \mathcal{M}_A) \mathcal{L}(d |\theta, \mathcal{M}_A)}{\mathcal{Z}(d| \mathcal{M}_A)}\,,
\end{equation}    
$p(\theta | d,\mathcal{M}_A)$ is the posterior distribution which gives the probability of all unknown parameters~$\theta$, given the experimental data $d$ within the model $\mathcal{M}_A$. $\mathcal{L}(d | \theta , \mathcal{M}_A)$ is the likelihood function. $\pi (\theta | \mathcal{M}_A)$ is the prior assuming $\mathcal{M}_A$. $\mathcal{Z}(d | \mathcal{M}_A)$ is the evidence which is the integral over the full set of parameters $\theta$ of the product of the likelihood and the prior:
\begin{equation}
\mathcal{Z}(d | \mathcal{M}_A) = \int d\theta \mathcal{L}(d | \theta, \mathcal{M}_A) \, \pi (\theta | \mathcal{M}_A) \,.
\label{evidence}
\end{equation}

Our goal is to study when BNS, BHNS and BBH mergers can be distinguished via the wave-forms of the emitted GWs. Bayesian analysis estimates the posterior probability density function for a given model $\mathcal{M}_A$. In order to compare two competing models, the odds number can be used~\cite{Thrane:2018qnx}:
\begin{equation}
\mathcal{B}_{AB} = \frac{p(\mathcal{M}_A |d)}{p(\mathcal{M}_B |d)} = \frac{p(\mathcal{M}_A)}{p(\mathcal{M}_B)} \frac{\mathcal{Z}_A}{\mathcal{Z}_B}\,,
\label{Bayes_factor}
\end{equation}
where $\mathcal{Z}_{A/B}$ is the evidence and $p(\mathcal{M}_{A/B})$ is the prior belief in model $A/B$. Thereafter, GW data are analyzed without assumption on the nature of the compact objects whichever their mass, and $p(\mathcal{M}_{A})/p(\mathcal{M}_{B})$ is fixed to one. $\mathcal{B}_{AB} = \mathcal{Z}_A / \mathcal{Z}_B$ is the Bayes factor between the two competing models $\mathcal{M}_A$ and $\mathcal{M}_B$. Jeffrey's scale~\cite{Jeffreys} gives an empirical calibration of the strength of evidence and if $\ln \mathcal{B}_{AB} > 5$, $\mathcal{M}_A$ is strongly preferred. Limits for strong evidence, moderate evidence and inconclusive are given in the Table~\ref{tab::bayes_factor}.

It is important to notice that a Bayes factor penalizes more complicated models with a large prior volume or a fine tuning. In other words, a particularly simple model giving a worse fit to the data can be preferred over a complicated model giving a better fit. This can be intuitively understood invoking the Occam's razor principle as explained in reference~\cite{MacKay}.

\begin{table}[h!]
\centering
\begin{tabular}{|c|c|c|}
\hline $|\ln \mathcal{B}_{AB}|$ & Probability & \\
\hline  $<1$    &   $<0.731$   & Inconclusive \\
\hline  $2.5$ & $0.924$ & Moderate evidence \\
\hline  $5$ & $0.993$ & Strong evidence \\
\hline
\end{tabular}
\caption{Jeffrey's scale to compare two competing models using the Bayes factor.}
\label{tab::bayes_factor}
\end{table}

%%%%%%%%%%%%%%%%%%%%%%%%%%%%%%%%%%%%%%%%%%%%%

\section{Model comparison with Bayes factor in injected data}

In this section we use {\tt IMRPhenomPv2}, {\tt IMRPhenomPv2\_NRTidal} and {\tt IMRPhenomNSBH} to describe BBH, BNS and BHNS wave-forms, respectively. The generated templates are used to analyze LIGO/Virgo data, and will constitute our model hypothesis to test against the simulated data. For this, we choose to perform a Bayesian analysis for each template and study its output in terms of the posterior distribution of the parameters and the signal-to-noise ratio. To determine which template better describes the data, we use the odds number introduced in the previous section. The power spectral density of the detector noise is taken to be the one of the advanced configurations for the O${}_4$ simulations\footnote{https://dcc.ligo.org/LIGO-T2000012/public}. As the nature of the compact objects is assumed unknown, three analyses are performed, one with the BBH system, one with the BNS system and finally one with the BHNS system. Then, we compute the Bayes factor between these models in order to find which one describes best the simulated data, with a decisive choice whenever the Bayes factor is higher than~$5$.  The results are computed using {\tt pBilby}~\cite{Ashton:2018jfp, Romero-Shaw:2020owr,Smith:2019ucc} with nested sampling, as introduced by Skilling \cite{Skilling:2006gxv}.

\subsection{Injected wave-form}
\label{sec:injected_parameters}
The first step in this study consists in generating a gravitational wave-form given by a template and adding it to the design Gaussian noise of advanced detector configurations as reported in~\cite{LIGOScientific:2019hgc}. The GW depends on $15$ parameters: chirp mass $\mathcal{M}$, mass ratio $q$, two angular momenta $\vec{S}_{1,2}$ (for aligned spin systems, only $\chi_i = \vec{S}_i.\hat{\vec{L}}/m^2_i$ are needed, where $\hat{\vec{L}}$ is the normalized orbital momentum), 2-dimensional sky localization (right ascension $ra$ and declination $dec$), luminosity distance $d$, two angles for the orbital plane (inclination angle $\iota$ and the polarization angle $\psi$), coalescence time $t$ and phase of coalescence $\phi$. For a BNS merger, $17$ parameters must be used because the two tidal deformabilities $\Lambda_1$ and $\Lambda_2$ must be added to the list.

The location of the source in the sky is chosen to maximize the signal for the most sensitive detector (optimal configuration). The PSD for the Virgo detector, which determines the sensitivity of the detector, is chosen for the advanced configuration, shown in purple in Figure~\ref{fig:noise}. For the Hanford and Livingston detectors, the PSD has been represented in orange and is the same for both. The source location is taken to maximize the signal for the Hanford detector. For a coalescence at a GPS time of $1.2\times 10^{9}$~s, this gives $5.49$~rad for right ascension and $0.81$~rad for declination. The inclination and polarization are chosen at $0$~rad. For a physical luminosity distance of $400$~Mpc, the effective distance for these parameters gives $400$~Mpc for Hanford, $449$~Mpc for Livingston, and $1867$~Mpc for Virgo; the antenna pattern of the emission is maximal for Hanford as expected. Unfortunately, the Virgo detector is almost blind. A simple change in declination improves the average effective distance of the three detectors. For example, a declination of $1.49$ rad instead of $0. 81$ rad gives an effective distance of $512$~Mpc for Hanford, $626$~Mpc for Livingston and of $590$~Mpc for Virgo. However, this change in declination does not improve the signal/noise Bayes factor: $\mathcal{B}_{SN} = \mathcal{Z}_S / \mathcal{Z}_N$ where $\mathcal{Z}_N$ is the noise evidence sometimes called null likelihood.  

The spin of neutron stars in a binary system is generally extremely weak. Well before the inspiral phase detected by gravitational wave detectors, the spin of neutron stars is suppressed by electromagnetic interactions. On the contrary, standard and primordial black holes, which do not have this suppression mechanism, can have larger spins during the merger, and a primordial black hole can even have a spin very close to that of an extreme Kerr black hole. If a compact object is detected with a large spin, there is a strong chance that it is a black hole and if the spin is almost equal to $1$: a primordial black hole. As for the mass criterion, we propose here to study the distinction between BH and NS only with the Bayes factor without adding assumptions in the prior and we will restrict ourselves to compact objects injected without spin. A priori, the presence of a spinning BH in the coalescence will not increase the odds number to distinguish two models, because it is always possible to assume a spinning NS. 

In the following, we will consider an injected signal with the optimal configuration for Hanford as described in above. Compact objects are chosen without spin and with masses equal to $m_1=1.74$~$M_{\odot}$ and $m_2=1.57$~$M_{\odot}$ i.e. a chirp mass of $\mathcal{M}=1.44$~$M_{\odot}$ with $q=0.9$ and $\chi_i=0$. For an injected BBH data created by {\tt IMRPhenomPv2} routine, only the luminosity distance varies between the different injections. For BNS data created by the {\tt IMRPhenomPv2\_NRTidal} routine,  the luminosity distance and the tidal deformabilities will be varied.

\begin{table}[t!]
\centering
\begin{tabular}{c|c|c|l|}
%  \cline{2-4}
%  & \multicolumn{3}{c|}{Bayesian analyses} \\
  \cline{2-4}
   & Name of models & Number & \hspace{2cm}Sampling parameters \\
\cline{2-4}

\hline 
\multicolumn{1}{|c|}{\multirow{5}{*}{\rotatebox{90}{Restricted space}}}   &  BBH W/O Spin & $2$ & $\mathcal{M}$, $q$\\
\cline{2-4}
\multicolumn{1}{|c|}{ }  &  BNS W/O Spin & $4$  & $\mathcal{M}$, $q$, $\Lambda_1$, $\Lambda_2$\\
 \cline{2-4}
\multicolumn{1}{|c|}{ }  &  BBH 1D Spin & $4$  & $\mathcal{M}$, $q$, $\chi_1$, $\chi_2$ (aligned spin)\\
 \cline{2-4}
\multicolumn{1}{|c|}{ }  &  BBH 3D Spin & $8$ & $\mathcal{M}$, $q$, $a_1$, $a_2$, $\theta_1$, $\theta_2$, $\phi_{JL}$, $\phi_{12}$ (full spin description)\\
 \cline{2-4}
\multicolumn{1}{|c|}{ }  &  BNS W/O Spin with $\Lambda_i$ fixed & $2$ & $\mathcal{M}$, $q$ ($\Lambda_1$ and $\Lambda_2$ are fixed at $0$)\\
\hline
\hline
\multicolumn{1}{|c|}{\multirow{4}{*}{\rotatebox{90}{Full space}}}   &  BBH & $11$ & $\mathcal{M}$, $q$, $\chi_1$, $\chi_2$, $ra$, $dec$, $d$, $\theta_{JN}$, $\psi$, $t$, $\phi$ \\
 \cline{2-4}
\multicolumn{1}{|c|}{ }  &  BHNS & $12$ & $\mathcal{M}$, $q$, $\chi_1$, $\chi_2$, $\Lambda_2$, $ra$, $dec$, $d$, $\theta_{JN}$, $\psi$, $t$, $\phi$\\
 \cline{2-4}
\multicolumn{1}{|c|}{ }  &  BNS & $13$  & $\mathcal{M}$, $q$, $\chi_1$, $\chi_2$, $\Lambda_1$, $\Lambda_2$, $ra$, $dec$, $d$, $\theta_{JN}$, $\psi$, $t$, $\phi$\\
 \cline{2-4}
\multicolumn{1}{|c|}{ }  &  BNS with $\Lambda_i$ fixed & $11$ & $\mathcal{M}$, $q$, $\chi_1$, $\chi_2$, $ra$, $dec$, $d$, $\theta_{JN}$, $\psi$, $t$, $\phi$ ($\Lambda_1=\Lambda_2=0$)\\
\hline
\end{tabular}
\caption{Sampling parameters used for the different Bayesian analyses. The parameters not mentioned in sampling parameters are set to the value used in the injected data.}
\label{tab::Sampling_params}
\end{table}

\subsection{Black hole mergers interpreted as neutron star mergers}
\begin{figure}[!ht]
\centering
\includegraphics[width=12cm]{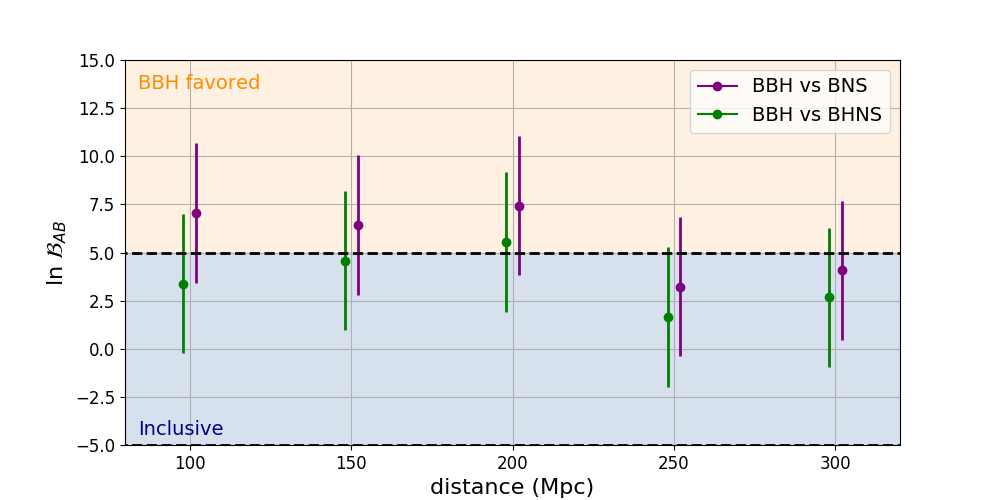}
\caption{Bayes factor for different models as a function of the luminosity distance for the optimal configuration. The chirp mass of BBH injected data is chosen to be $1.44$~$M_{\odot}$ with a mass ratio of $0.9$ and the sampling parameters for BBH, BHNS and BNS are given in Table~\ref{tab::Sampling_params}. A point above the dashed black lines means that the BBH hypothesis is strongly favoured. In the blue region labelled "Inclusive" it is not possible to strongly prefer one model over the other. 
\label{fig:full}}
\end{figure} 
The Bayes factor is calculated between BBH/BHNS models and BBH/BNS models for an injected BBH signal of $128$ seconds as described in the previous section. The sampling parameters used for the Bayesian analysis, performed using parallel bilby are given in Table~\ref{tab::Sampling_params}. The full set of parameters is searched: there are $11$ parameters for BBH, $12$ parameters for BHNS and $13$ parameters for BNS. The Bayes factor depends on the template but also on the prior. To reduce the dependence on the prior, the same is used for a BBH and for a BNS, namely we use a uniform prior between $0.87M_{\odot}$ and $5M_{\odot}$ for the chirp mass and between $0.125$ and $1$ for the mass ratio with constraints $1M_{\odot}<m_1<5M_{\odot}$ and $1M_{\odot}<m_2<3M_{\odot}$ for all the models. We also use aligned spins in the low spin limit corresponding to $|\chi_i|<0.05$ for each model. Only the tidal deformability, modeling the matter for a NS, has no correspondent for a BH. In this case, a uniform prior between $0$ and $5000$ for $\Lambda$ is added for a NS.

Figure~\ref{fig:full} shows the variation of the Bayes factor as a function of the distance. In all cases the values of $\ln \mathcal{B}_{AB}$ represented by the dots are positive, meaning that the $\mathcal{M}_A$ model is always preferred. If a point is above the dashed horizontal line, the $\mathcal{M}_A$ model is \emph{strongly} better (cf. Table~\ref{tab::bayes_factor}). The $\mathcal{M}_A$ model corresponds to a BBH merger, i.e. the model which corresponds to the BBH injected signal, and the $\mathcal{M}_B$ model corresponds either to a BNS merger or to a BHNS merger. For a distance smaller than $200$~Mpc, the purple dots in Figure~\ref{fig:full} are in the orange region and the BBH model is preferred over the BNS model which is not really the case for the BHNS model (green dots). We can therefore conclude that there is at least one black hole in the source that emitted the gravitational wave. The BNS wave-form with almost zero tidal deformability fits very well with the BBH wave-form, so the Bayes factor will prefer the model with the least parameter and as expected, the purple dots are above the green dots and above zero. Each dot plotted in Figure~\ref{fig:full} is the result of one or two independent Bayesian analyses and is accompanied by a vertical bar corresponding to the possible variation of the result: this bar represents the uncertainty of the Bayes factor and its variation due to different realizations of the noise. This uncertainty has been estimated with eight simulations in the zone where a model is strongly favored ($d=150$~Mpc) and in the "Inclusive" zone ($d=300$~Mpc). Later in the manuscript (Figure ~\ref{fig:full2}), we show those simulation points. When the odds number is close to the limit of $5$ it is not always possible to discriminate between the BBH and BNS models even with a source at a distance of $100$~Mpc.   

In this section and in the following we have assumed the source as oriented favorably for detection and the results are therefore rather optimistic. Moreover, we have considered a Gaussian noise for the detectors without considering terrestrial noise due for instance to meteorological conditions or human activity which is an ideal case. On the other hand, the limit can slightly improve with the increase of the chirp mass.

%%%%%%%%%%%%%%%%%%%%%%%%%%%%%%%%%%%%%%%%%%%%%

\subsection{Neutron star mergers interpreted as black hole mergers}

In this section, we consider the case where the injected data are generated by mergers of BNS and we try to determine whether it is possible to confuse a BNS merger with nonzero tidal deformability with a BBH merger. We inject the GWs generated by mergers of BNS with different tidal deformabilities and zero spin at different distances and study the Bayes factor between BNS and BBH.

\subsubsection{Study on spin in restricted parameter space}
\begin{figure}[!ht]
\centering
\includegraphics[width=12cm]{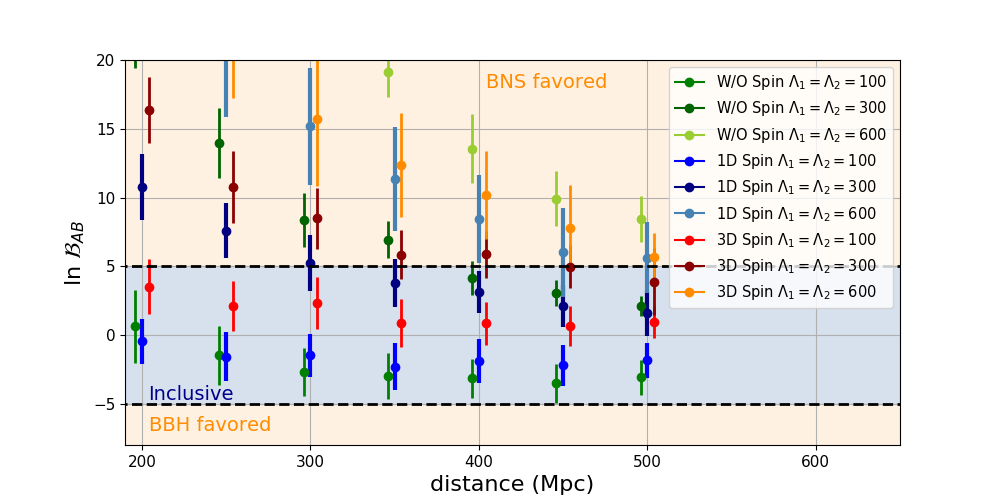}
\includegraphics[width=12cm]{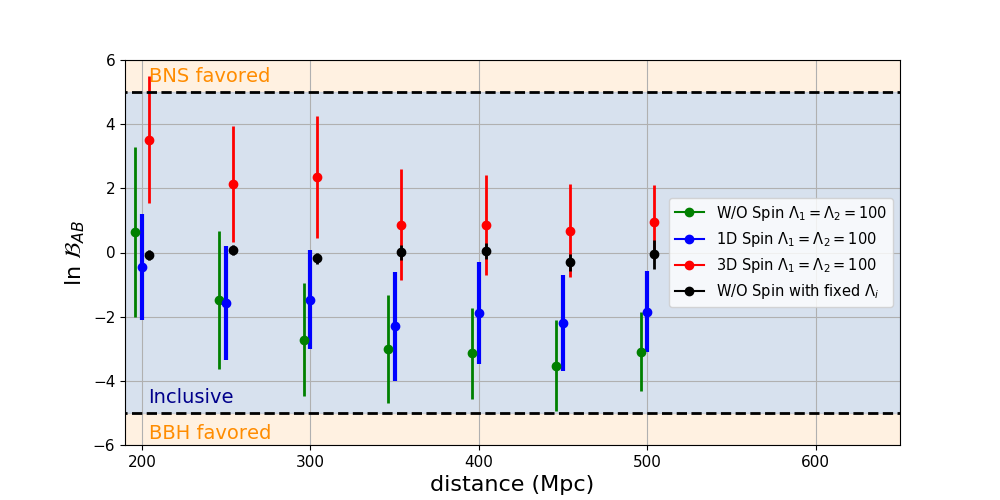}
\caption{Top panel: Bayes factor in restricted space between the model called "BNS without Spin" and different models of BBH as a function of the luminosity distance. The sampling parameters corresponding to the different models are given in Table~\ref{tab::Sampling_params} and the injected data correspond to BNS mergers in optimal configuration and with zero spins. Each point represents the mean value and the vertical bar the standard deviation for $4$ to $6$ simulations with different noise realizations. A dot located in the orange regions means that one of the two analyses is strongly preferred by the data. Bottom panel: zoom on the results for an injection with $\Lambda_1=\Lambda_2=100$. The black points are added and correspond to the Bayesian analysis performed with "BNS without Spin with $\Lambda_i$ fixed" and with "BBH without Spin" (see Table~\ref{tab::Sampling_params}).\label{fig:spin}}
\end{figure} 

First, we focus on the impact of the BH spins on the Bayes factor. In order to reduce the computational cost, only the physical parameters of the objects are searched for, using a Bayesian analysis. The sampling parameters for the different models considered are given in Table~\ref{tab::Sampling_params} and the results are shown in Figure~\ref{fig:spin}. In the top panel, the Bayes factor is calculated between the "BNS without Spin" model and a BBH model with different spin descriptions defined in Table~\ref{tab::Sampling_params}. For data injected with $\Lambda_1=\Lambda_2=600$ or with $\Lambda_1=\Lambda_2=300$, the "BNS without Spin" model is always preferred regardless of the BBH models. For data injected with $\Lambda_1=\Lambda_2=600$, the most favored case is the "BBH without Spin" model. Not taking spin into account in the BBH analysis, a BBH is unable to mimic a BNS and the posterior distribution of the mass ratio is not compatible with the injected value of $0.9$. The spin description allows for an improved signal fit with the BBH model due to the flexibility provided by a larger parameter space. On the other hand, the Bayes factor will not automatically be better because it penalizes models with a larger number of parameters, as shown by the blue and red dots, and using a 3D spin description is worse than a 1D description within the BBH model. For data injected with $\Lambda_1=\Lambda_2=100$, some values of $\ln \mathcal{B}_{AB}$ slightly prefer the BBH model in spite of the fact that the signal was created with a BNS model. The bottom panel of Figure~\ref{fig:spin} shows a new Bayesian analysis with the model "BNS without Spin with $\Lambda_i$ fixed" represented with black dots. In this case, the BBH model is no longer preferred compared to the BNS model.

\subsubsection{Full space parameters}
\begin{figure}[!ht]
\centering
\includegraphics[width=12cm]{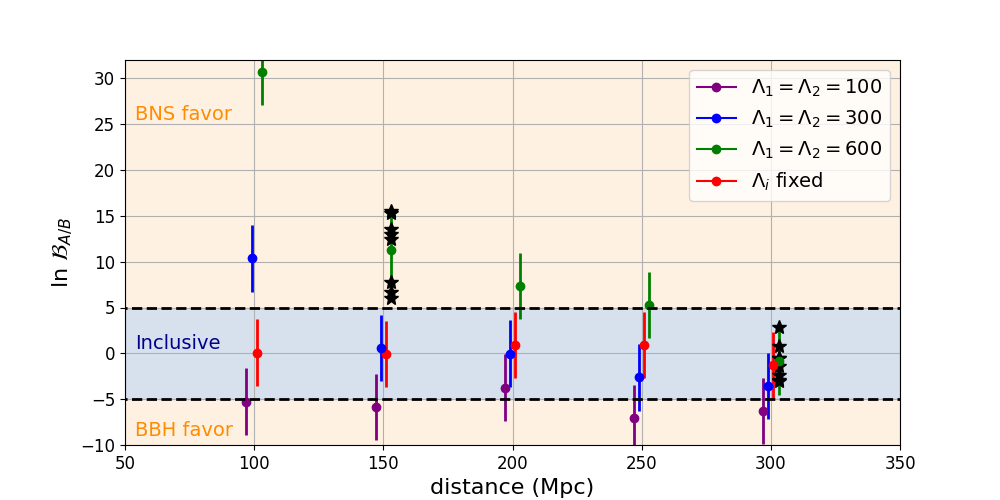}
\caption{Bayes factor in full space between BNS and BBH models as a function of the luminosity distance for different tidal deformabilities. The sampling parameters corresponding to the different models are given in Table~\ref{tab::Sampling_params} and the injected data are the same as in Figure~\ref{fig:spin}. A point in the orange regions means that one of the two models is preferred, whereas a point in the blue region corresponds to an inconclusive discrimination of the best model. The red points are obtained for the same injected data as for the purple points ($\Lambda_1=\Lambda_2=100$) but with the analysis called "BNS with $\Lambda_i$ fixed" which sets the tidal deformabilities to zero. The variation of the results using different realizations of the noise is represented by the black stars, which were used to draw the vertical bars.\label{fig:full2}}
\end{figure} 
In this section, the complete parameter space is probed via a Bayesian analysis (see Table~\ref{tab::Sampling_params}) for the same injected data as in Figure~\ref{fig:spin}. The Bayes factor between BNS and BBH is shown in Figure~\ref{fig:full2}. As expected, the larger the tidal deformability, the clearer the distinction and the less a BBH merger can mimic a BNS merger. We can also notice the strong impact of the distance from the source on the results. When $d=100$ Mpc, the tidal deformability is clearly reconstructed if $\Lambda_1=\Lambda_2\geq 300$. For a distance around $200$ Mpc, the Bayes factor is not systematically in the orange region even for a strong tidal deformability and for a distance of $300$ Mpc, it is no longer possible to discriminate the nature of the compact objects. 

When the injected signal is a BBH (see Figure~\ref{fig:full}), the Bayesian analysis never prefers the BNS model even if it is not possible to conclude that it is a BBH because the Bayes factor does not reach the strong evidence limit. The purple dots in Figure~\ref{fig:full2}, which correspond to a BNS signal with $\Lambda_1=\Lambda_2=100$, seem to suggest that the BBH merger is preferred over the BNS one. This preference is just a reminder that the BNS model is more complex than the BBH model because it has two additional parameters. This is demonstrated by the red dots which are consistent with a zero Bayes factor when the Bayesian analysis for BNS is performed by setting $\Lambda_1=\Lambda_2=0$ ("BNS with $\Lambda_i$ fixed" model).

As shown in Figure~\ref{fig:full2}, considering the PSD of the advanced detector configurations, the Bayes factor does not distinguish between the BNS model and the BBH model when the source is at $300$~Mpc even in the case where the tidal deformabilities are large ($\Lambda_1=\Lambda_2=600$). A significant improvement is expected with Einstein Telescope~\cite{Kroker:2015pmg}. To have an idea of this improvement, a BNS signal at $300$~Mpc is injected in the Hanford detector with the PSD~\cite{Hild:2008ng} of Einstein Telescope and in the Livingstone detector with the PSD of the advanced detector configuration. For a BNS signal with $\Lambda_1=\Lambda_2=600$, the Bayes factor is about $90$ and the BNS model is strongly favored compared to the BBH model. The distinction is even easier than with the previously considered signal which was injected at $100$~Mpc (see Figure~\ref{fig:full2}). For a BNS signal with $\Lambda_1=\Lambda_2=100$, the Bayes factor is now well above the $-5$ limit while staying in the region labelled "Inclusive".

%%%%%%%%%%%%%%%%%%%%%%%%%%%%%%%%%%%%%%%%%%%%%

\section{Conclusion}

The nature of the compact $2.6$~$M_{\odot}$ object detected in the GW190814 event is unknown: is it a neutron star, a black hole, a new type of compact object? Motivated by this event, we studied how a gravitational wave-form of a BBH merger can be distinguished from a BNS merger. Normally, the distinction is made by considering that there is a mass gap between black holes and neutron stars: a neutron star has a mass lower than $2.2$~$M_{\odot}$ and a black hole higher than $5$~$M_{\odot}$. The problem with the observed $2.6$~$M_{\odot}$ object is that it does not fit into this classification. Moreover, if primordial black holes exist, they may have masses similar to neutron stars. Thus, we reconsidered the distinction between a black hole and a neutron star by considering only the gravitational waveform without priors on the masses.

As a first step, we compared a wave-form from a BBH template, a BNS template and a BHNS template by using the match function defined in Eq.~\eqref{eq:match}. In the case where the two objects are described by the same physical parameters: same spin, same mass, the match between the different gravitational wave-forms is close to $1$. The parameter that allows us to distinguish them is the tidal deformability of the neutron star. In absence of tidal deformability, it will be impossible to have a difference. We have seen that even with a large tidal deformability, the observation of the gravitational wave of a merger at $400$~Mpc will not identify a difference between the templates. For a highly asymmetric BHNS merger, the tidal deformability of the companion is not a distinguishing parameter with a merger.

During a detection, the parameters of a fusion are determined by a Bayesian analysis. Such an analysis must use the template corresponding to the nature of the compact objects. A selection criterion for the nature of the fusion is given by the Bayes factor. In the case where two models $A$ and $B$ are in competition: a BBH fusion for $A$ and a BNS fusion for $B$ for example, the quantity $\ln \mathcal{B}_{AB}$ determines the best model. If $\mathcal{B}_{AB}>5$, model $A$ is strongly favored compared to $B$ and conversely if the ratio is less than $-5$. If the ratio is between these two values, the data can be explained by both models.We have therefore studied this ratio for data injected by considering a fusion with $1.44$~$M_{\odot}$ for the chirp mass. This factor depends on the prior volume and penalizes the most complex systems with a larger number of parameters. For BBH injected data, the Bayes factors of the BBH models are always better compared to the others. In such a merger, the wave-form is very well reproduced by a BNS model with neutron stars having a tidal deformability equal to zero but because of the tidal deformability, the BNS model is more complex and therefore worse than the BBH model. For injected BNS data with $\Lambda_1=\Lambda_2=600$, the Bayes factors of the BNS models are largely favored over BBH. On the other hand, when $\Lambda_1=\Lambda_2=100$, we find that the BBH model is better than BNS when the data is made with BNS. A better fit does not mean a better Bayes factor given by the evidence. To get rid of this effect, it is possible to fix the tidal deformability of neutron stars at zero to have exactly the same prior volume. In this case, either the Bayes factor becomes zero and one cannot be sure of the nature of the compact object or it remains negative and lower than $-5$ and the nature of the compact object becomes known. A very conservative limit to avoid preferring BBH when a BNS signal is injected without redoing a Bayesian analysis and fixing the deformability would be to take a limit of $8$ instead of $5$. This limit of $8$ has been conventionally introduced in~\cite{Thrane:2018qnx}.

The detection of a merger at a distance larger than $250$~Mpc will not have a good enough signal to noise ratio to allow the determination of the nature of compact objects even if the merger is perfectly oriented for the Hanford detector. This limit can be improved if we consider a higher chirp mass for the merger but during a real detection, the noise is never ideal and the orientation of the merger not optimal. By considering Einstein Telescope, the comparison between two models is dramatically improved. This ability to discriminate between two models can be used to interpret experimental measurements of GWs and obtain incredible insights in General Relativity.  

\bibliography{biblio}

\end{document}